\documentclass[conference]{IEEEtran}
\usepackage{graphicx,times,amsmath,amssymb,multirow,caption,subcaption,color}
\usepackage[LGR,T1]{fontenc}
\usepackage[utf8]{inputenc}
\usepackage{array,textcomp,stackrel,url,mathtools,enumerate}
\usepackage{multirow}  
\usepackage{multicol}  
\usepackage{booktabs,helvet,courier,float,csquotes}
\usepackage{algorithm,algcompatible} 
\usepackage{csquotes}
\usepackage[colorinlistoftodos]{todonotes}
\usepackage{color}
\usepackage{color,soul}
\usepackage[inline]{enumitem}
\usepackage{cite}
\newcolumntype{x}[1]{>{\centering\arraybackslash\hspace{0pt}}p{#1}}
\makeatletter

\begin{document}
\title{Transparent Machine Education of Neural Networks for Swarm Shepherding Using Curriculum Design}

\author{\IEEEauthorblockN{Alexander Gee}
\IEEEauthorblockA{\textit{School of Engineering \& IT} \\
\textit{University of New South Wales}\\
Canberra, Australia\\
a.gee@unswalumni.com}
\and
\IEEEauthorblockN{Hussein Abbass}
\IEEEauthorblockA{\textit{School of Engineering \& IT} \\
\textit{University of New South Wales}\\
Canberra, Australia\\
h.abbass@adfa.edu.au}
}

\maketitle

\begin{abstract}
Swarm control is a difficult problem due to the need to guide a large number of agents simultaneously. We cast the problem as a shepherding problem, similar to biological dogs guiding a group of sheep towards a goal. The shepherd needs to deal with complex and dynamic environments and make decisions in order to direct the swarm from one location to another. In this paper, we design a novel curriculum to teach an artificial intelligence empowered agent to shepherd in the presence of the large state space associated with the shepherding problem and in a transparent manner. The results show that a properly designed curriculum could indeed enhance the speed of learning and the complexity of learnt behaviours.
\end{abstract}


\section{Introduction}\label{intro}

\IEEEPARstart{T}{he} shepherding problem involves externally influencing a swarm or a group of agents to guide them towards the Shepherd\textquoteright s goal. The problem is algorithmically difficult due to the complexity of its search space~\cite{lienPratt2009}. Biological Shepherds such as Dogs are naturally smart. When a shepherd is an uninhabited ground or aerial vehicle, the complexity of the shepherd\textquoteright s problem increases orders of magnitude such that a machine learning approach becomes necessary. An artificial intelligence empowered shepherd \textit{AI-empowered Shepherd} (AIES) is a smart agent for controlling a swarm.

The two challenges associated with the development of a neural network for shepherding are: (1) the search space is large and thus, obtaining a functioning network in complex scenarios is non-trivial, and (2) neural networks are black-box models that makes it difficult to understand the skills being learnt. These challenges call for  functional decomposition of the overall task to dissect the learning problem into chunks. Such an approach could be traced back to Elman\textquoteright s seminal paper, \textit{The Importance of Starting Small}~\cite{elman1993}, where he explored the relationship between cognitive and memory capacity limitations of early developmental stages (of children), the rate at which learning is able to take place, and whether this concept has an application in machine learning. 

Elman suggested that it was these initial limitations, which forced the learner to `start small' (i.e. learning simple concepts first before moving onto more complex/abstract concepts), that was responsible for the accelerated learning rate. His ideas were experimentally demonstrated by first limiting the inputs the learner received to simple ones, and increasing their complexity; and secondly by incrementally increasing the limited memory capacity of the learner while keeping the input constant, resulting in a learner that was able to learn a `semi-realistic artificial language' while reaching a level of competency that was not previously reachable without the constraints imposed by the incremental approach.

Elman\textquoteright s gave birth to a school of thoughts where theories for curriculum design in human learning became applicable to machine learning. Work in curriculum learning~\cite{bengio2009,elman1993} has demonstrated the learning efficiency benefits when applied to machine learning; and improved scalability of specialised data creation through expert demonstration~\cite{abbeelNg2004} and game play~\cite{barrington2012} has also been demonstrated. Bengio et al.~\cite{bengio2009} were then able to show significant improvements in speed of convergence to local minima, as well as the quality of the local minima obtained, by implementing `Curriculum Learning'  which is the concept describing the training strategy of organising concepts from simple to more complex ones before presenting them to a machine learner.  

The concept of a curriculum has been demonstrated further in~\cite{florensa2017} where the task of a robot aligning a gear onto an axle is taught, by training a reinforcement learner in \enquote{reverse.} This is achieved by initially causing start states to be close to the goal (i.e. gear oriented correctly above axle) and as it learns to achieve these small tasks, the start states are set increasingly far from the goal. 

Curriculum-based learning requires labelled data, a cumbersome task. Barrington et al.~\cite{barrington2012} proposed a model for game-powered machine learning which they claim combines the effectiveness of \emph{human computation} with the scalability of \emph{machine learning} through the use of games. They suggest that when there is a need for large data sets, manual tagging/labelling done by human experts or through crowdsourcing is both too costly and too time-consuming. Instead, they suggest online crowdsourcing in the form of a game, be used to provide the training set for a machine learning algorithm. 

The area of swarm control is properly one of the most difficult learning problems, which calls for a curriculum-based approach. Nevertheless, the literature lacks research in this area, which motivated the work in this paper.

In this paper, we propose a curriculum learning approach to design an AIES. A curriculum is designed hierarchically by structurally decomposing the task of shepherding into sub-behaviours and their required skills. The curriculum design is coupled with apprenticeship learning~\cite{abbeelNg2004,nguyen2018}, with each skill associated with a lesson learnt from human data.

By following a learner-centric approach, curriculum design could account for the intended machine learning algorithm and tailor elements of the design to the particularities of the algorithm.  Consequently, we adopt George\textquoteright s \textit{Classical Curriculum Design}~\cite{george2009}. 

We use an interactive simulation to generate labelled data for different skills required for shepherding. Different skills were associated with different scenarios, which allowed for the creation of the labelled dataset necessary for the AI learner to develop the required skills. We then aggregate these skills into an overall AIES for shepherding. We use a basic form of shepherding to demonstrate the proposed general methodology.

The remainder of this paper is organised as follows. The basic shepherding model used to create human demonstration data is covered in Section~\ref{Shepherding} followed by the proposed methodology, experiments, results, 
and conclusion in Sections~\ref{methodology},~\ref{experiments},~\ref{results},
and~\ref{conclusion}, respectively.

\section{Shepherding}\label{Shepherding}

\begin{tabular}{ll} 
$\Psi$      & The shepherd\textquoteright s output direction vector \\ 
$\Psi_D$    & Shepherd\textquoteright s output driving direction vector\\ 
$\Psi_C$    & Shepherd\textquoteright s output collecting direction vector \\ 
$P_F$       & Position of furthest agent from GCM \\ 
$P_G$       & Position of GCM (global centre of mass) \\ 
$P_C$       & Position of calculated collection point \\ 
$P_D$       & Position of calculated driving point \\
$P_T$       & Position of target \\ 
$P_S$       & Position of shepherd \\ 
$R$         & Magnitude of GCM to furthest agent distance \\
$N$         & Number of agents in cluster \\ 
$f(N)$      & Switching distance \\ 
$n$         & Current time step \\ 
$r_a$       & Agent-to-agent interaction distance\\ 
\end{tabular}

According to the model suggested by Str\"{o}mbom et al.~\cite{strombomMann2014}, shepherding brings together two specific behaviours. Firstly the shepherd responds to a dispersed herd by rounding up stray agents, or \textit{collecting}; and secondly it responds to an aggregated herd by pushing them towards the goal, or \textit{driving}. 

Given a shepherd position $P_S$, if the furthest agent has a distance $R$ from the global centre of mass (GCM) greater
than a distance $f(N)$, the shepherd moves in the direction of the collection point $P_C$; point directly behind the sheep (relative to the GCM) with greatest distance from GCM. Otherwise if all agents are within distance $f(N)$ from the GCM, the shepherd moves towards the driving point $P_D$; point directly behind the GCM relative to the goal location and at the outer boundary of the herd. The switching behaviour between collection and driving is represented as shown in Equation~\ref{eq:switchEquation}, where $\Psi(n)$ is the final shepherd direction vector in timestep $n$:

\begin{align}\label{eq:switchEquation}
        ||P_F(n) - P_G(n)|| <   f(N) \to \Psi(n) =  & \Psi_D(n);\nonumber\\
        \text{else } \Psi(n)                     =  & \Psi_C(n)
    \end{align}
    
Where the instantaneous shepherd driving direction vector $\Psi_D(n)$ can be modelled as shown in equation \ref{eq:driveVect}.
    \begin{align}\label{eq:driveVect}
        \Psi_D(n)   & = \frac{P_D(n) - P_S(n)}{||P_D(n) - P_S(n)||}\\
        P_D(n)      & = \frac{P_G(n) - P_T(n)}{||P_G(n) - P_T(n)||}r_a\sqrt{N} + P_G(n)
    \end{align}
    And similarly the instantaneous shepherd collection direction vector $\Psi_C(n)$ can be found as shown in Equation \ref{eq:collectVect}
    \begin{align}\label{eq:collectVect}
        \Psi_C(n)   & = \frac{P_C(n) - P_S(n)}{||P_C(n) - P_S(n)||}\\
        P_C(n)      & = \frac{P_F(n) - P_G(n)}{||P_F(n) - P_G(n)||}(R + r_a) + P_G(n)
    \end{align}
    
%

The sheep follows swarming rules. This is justified by the selfish herd theory~\cite{hamilton1971}, which shows that individual agents will form herds to reduce their own individual risk of predatory attack. 
The seminal swarming model introduced by Reynolds \cite{reynolds1987} offers the three basic rules to implement the selfish herd theory; these rules are separation, alignment, and cohesion. The separation rule is a collision avoidance rule to maintain the sheep dispersion and acts as a repulsive force between one sheep and another. The alignment force ensures that an agent aligns its velocity with its neighbours. However, in the Str\"{o}mbom model, this rule is replaced with an inertia so that a sheep aligns with its previous velocity vector. The cohesion rule groups sheep together by applying an attraction force from a sheep to the local centre of mass of its neighbourhood. In addition, Str\"{o}mbom adopts two extra rules to each sheep. One is a repulsion force from the shepherd and the second is an angular noise as a disturbance term to a sheep position.

\section{Methodology}\label{methodology}

\begin{figure}[phtb]
    \centering
\includegraphics[width=0.85\columnwidth,height=0.6\textheight]{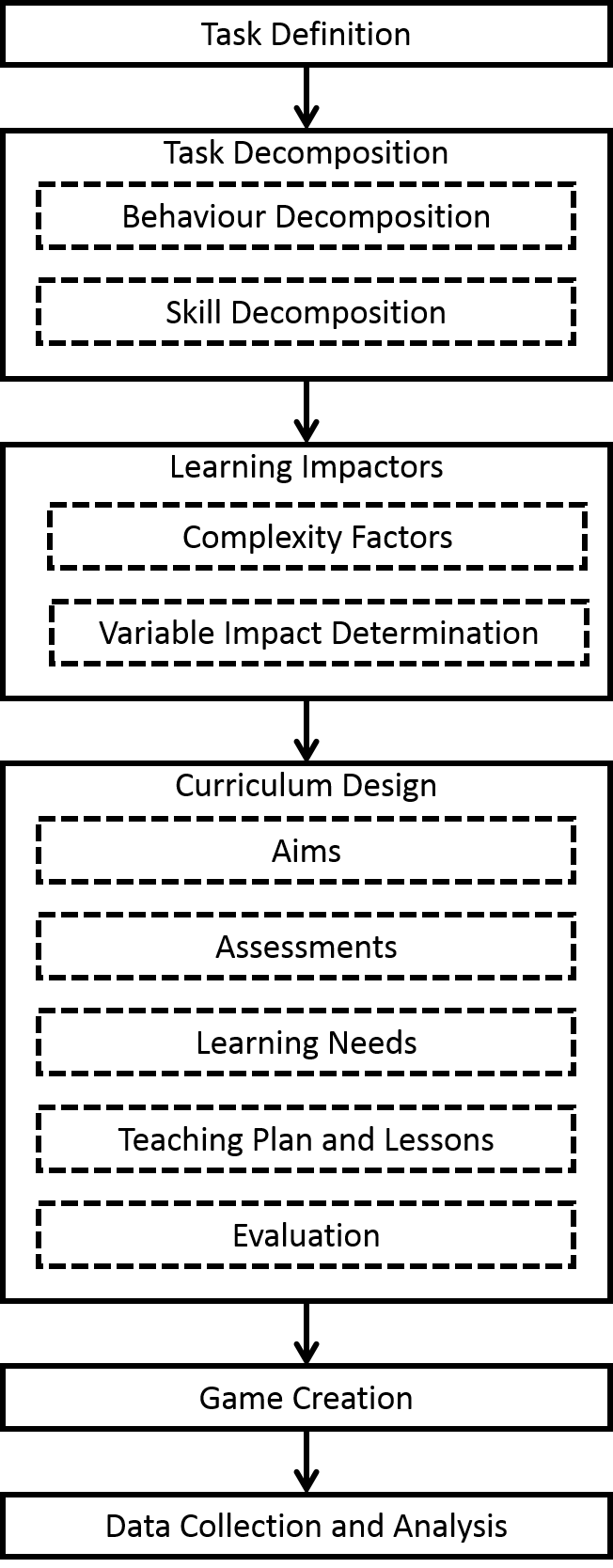}
    \caption{Six Stages of AI education}
    \label{fig:Methodology}
\end{figure}

In this section, we present our proposed curriculum design methodology to teach an AIES to shepherd a swarm of agents, while using an interactive simulation to generate demonstrations for the machine learner to learn from. The six-stage curriculum methodology is based on George\textquoteright s \textit{Classical Curriculum Design}~\cite{george2009} with modifications to make it suitable for a machine learner. In particular, we added five stages (1-3, 5 and 6). In George\textquoteright s methodology, the assumption is that there is a human who is an experienced curriculum designer and is experienced in the subject matter of the curriculum. We do not make this assumption here. Instead, we design the methodology for data engineers to be able to adopt it without the need to be immersed in curriculum design. The fourth and fifth stages are designed to create the lessons for the machine learner. The six stages are depicted in Figure~\ref{fig:Methodology}.

\subsection{Task definition}  The designer needs to be clear on what the task is and the desired solution. The AI focuses on solving the \enquote{how}, while the designer is focused on solving the \enquote{what}; that is, what behaviours are expected from the AI and what are the tasks entrusted to the AI? The task definition statement defines the requirements for the AI. In this paper, the task of the AI is described as follows:
\begin{quote}
\textit{With a single shepherd, from any starting location, influence a swarm of 50-200 agents towards a single stationary goal predefined at a location in an obstacle free environment.} 
\end{quote}

\subsection{Task Decomposition}
Two types of interdependent forms of task decompositions need to be resolved. The first is behavioural decomposition. A behaviour defines the displayed pattern when an agent acts. The second is skill decomposition. A skill defines the know-how required to produce the right actions. Each high level behaviour requires one or more skills strung together in various patterns to express the target behaviour.

\subsubsection{Behavioural decomposition} 
Inspired by Str\"{o}mbom et al\textquoteright s model~\cite{strombomMann2014}, shepherding requires two basic behavioural building blocks and the switching between them, namely, \textit{collecting} and \textit{driving}.
    
In the scripted model of Str\"{o}mbom, switching occurs based on the distance of the furthest agent from the GCM. The shepherd has to decide whether it is firstly too close to the sheep, in which case it will stop moving; and secondly decide whether the herd is collected to adopt the driving behaviour, otherwise it adopts a collecting behaviour.

Based on our mathematical representation of Str\"{o}mbom\textquoteright s \cite{strombomMann2014} model described earlier, an agent learning all behaviours at once to generate the output vector $\Psi$ could be described mathematically as:  

\begin{align}\label{eq:shepherdEquation}
    \Psi    & = F(P_F, P_G, P_S, P_T, r_a, N)
\end{align}

A curriculum-based agent will have the two collecting and driving behaviours denoted by $\Psi_C$ and $\Psi_D$, respectively and represented as

\begin{align}
    \Psi_C = F(P_S, P_F, P_G, r_a). 
\end{align}

\begin{align}
    \Psi_D = F(P_S, P_G, P_T, r_a, N).
\end{align}

The curriculum-based agent will generate the output vector $\Psi$ by switching between the two behaviours above as follows:

\begin{align}\label{eq:shepherdEquationC}
    \Psi    & = F(S, \Psi_C, \Psi_D)
\end{align}

where, $S$ is the switching meta-behaviour defined as 
\begin{align}
S   & = F(P_F, P_G, r_a, N)
\end{align} 

In our case, this switching behaviour is pre-scripted in our curriculum based agent using the same condition by Str\"{o}mbom of whether the furthest agent distance to the GCM was less than or greater than $f(N)$, whereas the non-curriculum agent needed to learn this behaviour within its single neural network. The simplicity of the switch in this case did not require an independent neural network. A more complex switch, however, would be learned in the same way we learn the individual behaviours.

\subsubsection{Skill decomposition}  
Skills are generally broken up into discrete, serial and continuous skills\cite{website:skills2018}. Shepherding is a serial skill comprising a number of discrete skills put together in a variety of patterns. To learn to shepherd, firstly a solid grasp of the fundamental discrete skills is required. Secondly once these have been learned, the patterns with which to string these together must be learned and understood. Quantifying the complexity of the discrete and serial skills should occur at this stage as this will act as a guide for lower level curriculum design decisions. Each distinct act is dependent on various low level skills already being applied to attempt the learning of the acts, which in turn are required to learn the higher level collecting and driving behaviours.
    
\textit{Shepherd-agent interactions:} The first of these lowest level skills is an understanding of the shepherd-agent interactions that take place, i.e. an answer to the fundamental question, what happens when the shepherd moves closer to an agent? To go about defining $P_C$ and $P_D$, understanding that the sheep will be repelled from the shepherd is crucial.
    
\textit{Movement:} A necessary lower level skill is the ability of the shepherd-agent to move within the environment. This may sound trivial, but in practice this translates into a very complex skill where the agent needs to actuate on the environment and decides when and when not to actuate. 

\textit{Estimating the collection and driving points:} The shepherd needs to know where it needs to go. Movement assists it to reach the destination, but estimating the destination is a skill in its own right.

\subsection{Learning Impactors}

\subsubsection{Complexity Factors} 
Another important factor in the curriculum concerns the quantification of the level of difficulties associated with a task. Task complexity raises two questions.

\begin{figure*}[ht]
\centering
\begin{subfigure}{0.3\textwidth}
    \centering
\includegraphics[width=0.95\columnwidth,height=0.12\textheight]{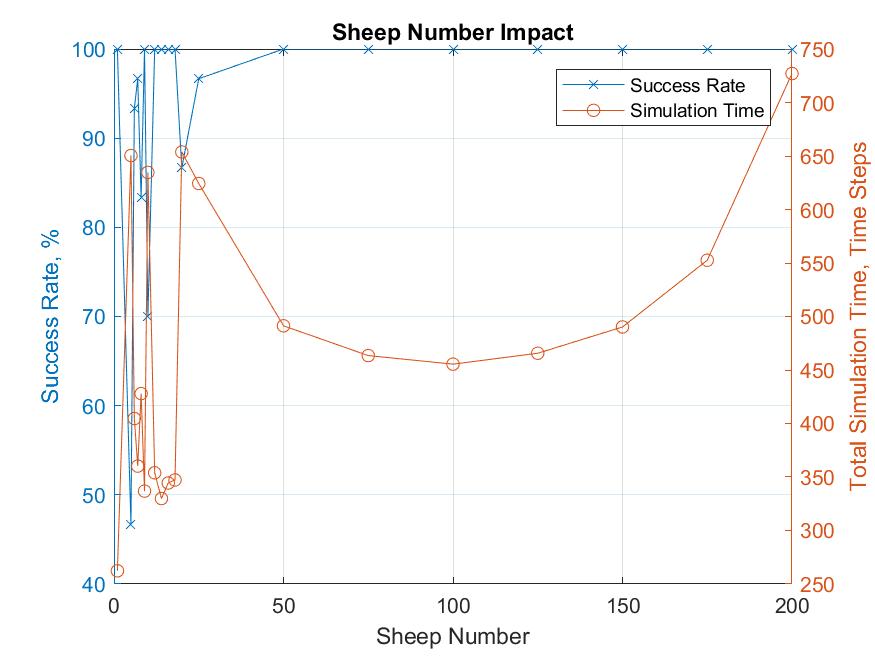}
    \caption{Sheep Number}
    \label{fig:sheepNumPlot}
\end{subfigure}
\begin{subfigure}{0.3\textwidth}
    \centering
\includegraphics[width=0.95\columnwidth,height=0.12\textheight]{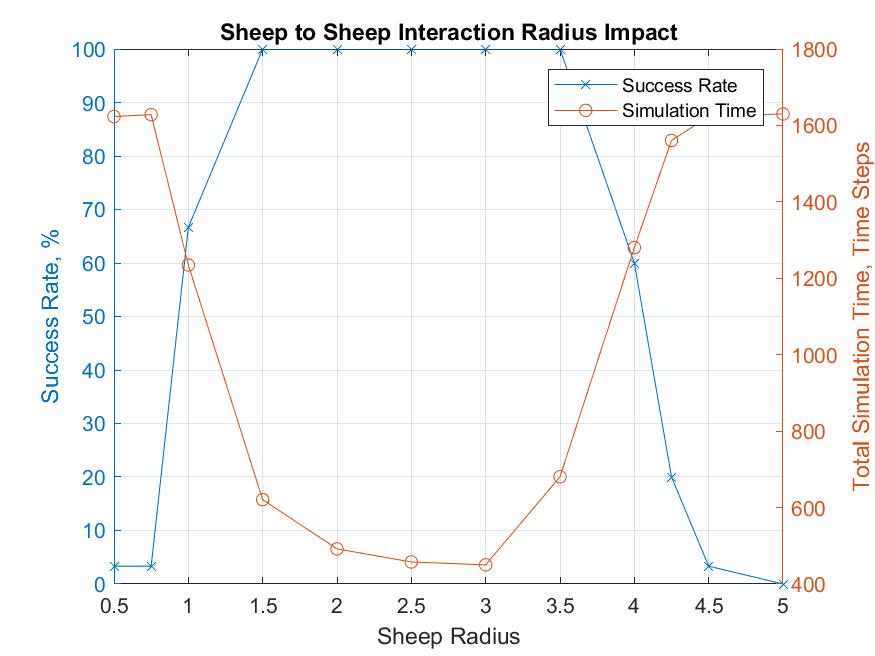}
    \caption{Sheep Radius}
    \label{fig:sheepRadPlot}
\end{subfigure}
\begin{subfigure}{0.3\textwidth}
    \centering
\includegraphics[width=0.95\columnwidth,height=0.12\textheight]{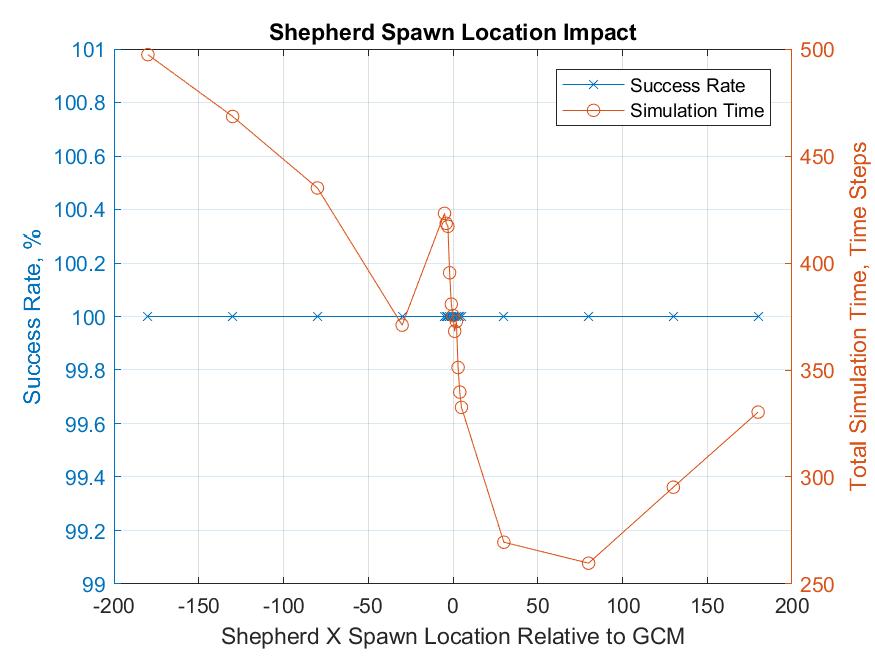}
    \caption{Shepherd Spawn X Location (relative to GCM)}
    \label{fig:shepSpawnLocPlot}
\end{subfigure}
\begin{subfigure}{0.3\textwidth}
    \centering
\includegraphics[width=0.95\columnwidth,height=0.12\textheight]{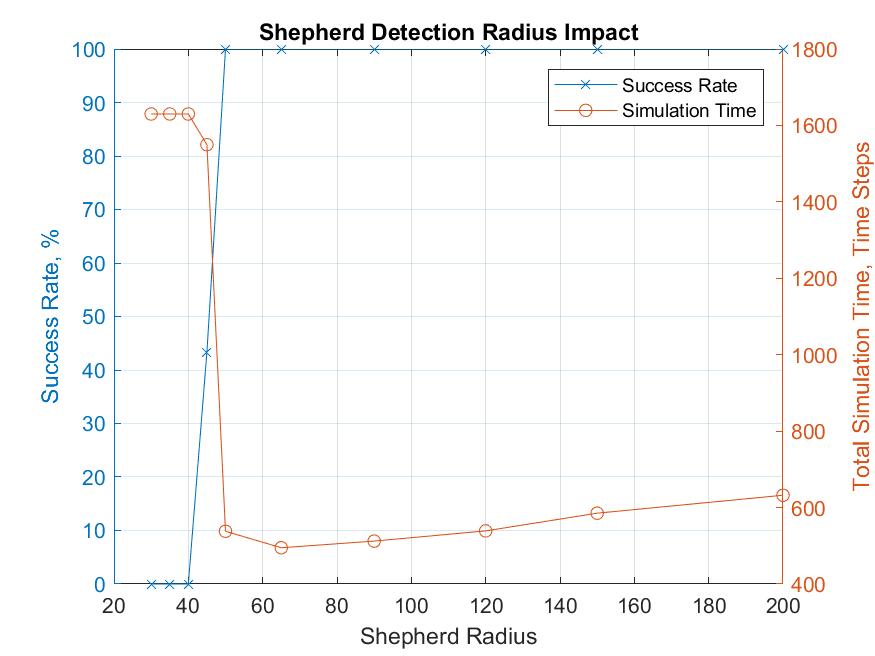}
    \caption{Shepherd Radius}
    \label{fig:shepRadPlot}
\end{subfigure}
\begin{subfigure}{0.3\textwidth}
    \centering
\includegraphics[width=0.95\columnwidth,height=0.12\textheight]{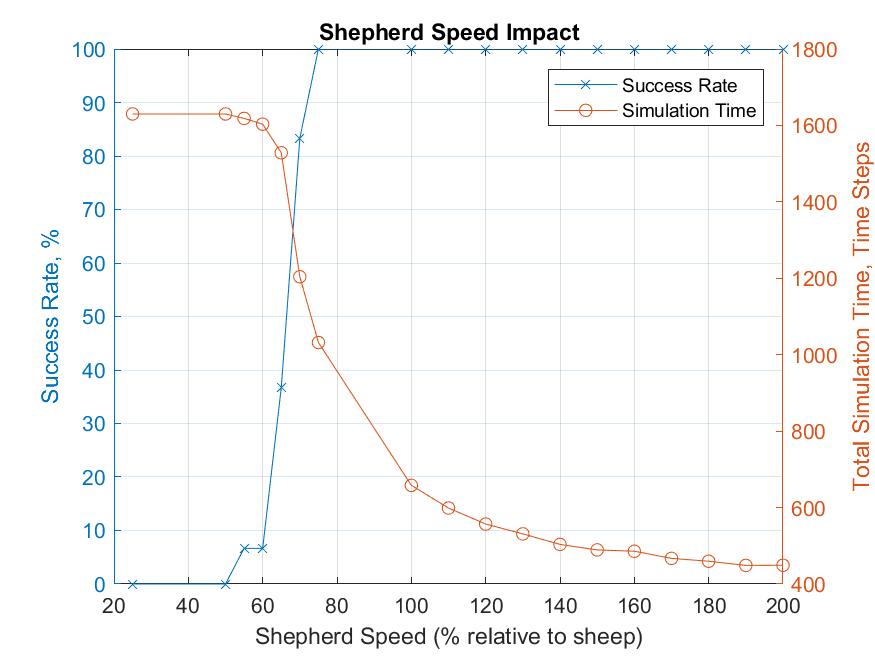}
    \caption{Shepherd Speed}
    \label{fig:shepSpeedPlot}
\end{subfigure}
\caption{Variable Impacts on Difficulty Simulation Results}
\label{fig:difficultyVariablePlots}
    \end{figure*}

Firstly what makes it difficult to learn that skill to begin with?  Secondly what factors affect the difficulty of completing the task; that is, the identification of the set of factors that control task complexity? The first question calls for metrics to assess task complexity while the second calls for tracing the causes for complexity. To address both questions, a series of experiments were conducted to determine the effects of changing various variables on difficulty with the results shown in Figure~\ref{fig:difficultyVariablePlots}. Two metrics to indicate complex tasks are used: Success Rate and Completion Time. The former is calculated as the ratio of successful trials to the total number of trials. The latter is calculated as the total simulation time of each simulation.

These are complementary metrics and neither of them could independently capture task-difficulty. For example, a 100\% success rate hides the variance associated with the time taken to achieve this success. The time to complete a task, if used alone, could be misleading as it is natural that a simple increase in distance between the initial position of the shepherd and the sheep will cause an increase in task completion time, assuming everything else is constant. Nevertheless, this increase in time is deceptive as it is merely a scaling effect.

\begin{figure}[ht]
    \centering
\includegraphics[width=0.9\columnwidth]{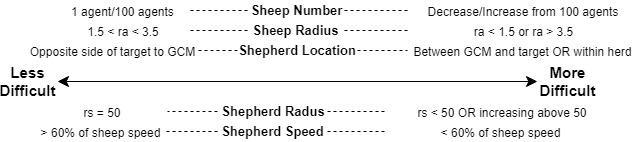}
    \caption{Difficulty Factor Summary}
    \label{fig:difficultyDiagram}
\end{figure}

We have identified five factors based on Str\"{o}mbom\textquoteright s model that impact the model, and therefore are expected to impact the two metrics above. These are the number of sheep, the radius where the sheep gets initialised, the number of locations a sheep spawn relative to the global centre of mass, the shepherd radius and relative speed of shepherd to sheep. The results of complexity are shown in Figure~\ref{fig:difficultyDiagram}.

\subsubsection{Variable Impact Determination} 
For sake of competence of the methodology, we need to ask if there is a particular weighting scheme that should be used to combine the different factors impacting task complexity to form an overall complexity index. In the case of humans, this is normally done through human experiments with questionnaires associated with workload. In the case of a computer system, we will assume in this paper that these factors are weighted equally.

\subsection{Curriculum Design} 
The previous steps conclude with an understanding of what makes a task complex, which represents the starting point for designing a curriculum to educate the AI. The curriculum design step consists of the a number of sub-steps presented and discussed below. 

\subsubsection{Aims} 
From the task definition defined in step 1, the main aim of the learner is to learn how \enquote{to shepherd 50-200 agents from any starting location, to any goal location.} This aim offers the point of reference to measure the complexity of the task that has been discussed in the previous sub-section.
    
\subsubsection{Assessments} 
The chosen assessment is designed around the above desired learning aims. The \textbf{purpose} of the assessment is to confirm that the necessary skills have been learned before moving on to the next lesson/level. The \textbf{context} of the assessment is that the learning and assessment environment are aligned, with assessment applying the trained skill with untrained variables. The assessments should directly assess the main \textbf{learner aims} pulled from the task statement. 

Two types of assessments need to be considered: summative and formative. For summative assessments, we run 100 simulations of the open environment with each AEIS, and compare their ability to drive, collect and shepherd overall. For formative assessments, we run 50 simulations of the open environment with the AEIS at increasing sample size to gain understanding of the learning rate of each AEIS; that is, generalisation ability of the learner as a demonstration that the learner truly learnt, not just memorised, the skills. The lower number of simulation runs in formative assessments is due to their narrower search space compared to summative assessments. 

Summative assessment and formative assessments are counted as successful if the minimum GCM to goal distance, and average furthest agent to GCM distance achieved by the curriculum based AIES is less, and the success rate is higher than those achieved by the non-curriculum based AIES.

\subsubsection{Learning needs} 
Selecting the correct data that fully represents the skills and behaviours to be learned is crucial. Data must be presented in a way that encourages the learner to determine patterns from specific scenarios, that can be applied to the general case or risk the neural network developing weightings biased to the training scenarios. For this reason as much of the data as possible is presented relative to the shepherd. Noting Equation~\ref{eq:shepherdEquation}, we know that the desired output direction is some function $\Psi = F(P_F, P_G, P_S, P_T, r_a, N)$; hence these independent variables are provided as training samples. The learner has one output representing the velocity of the agent. The inputs are defined below:

\begin{enumerate}
\item Shepherd to goal vector
\item Shepherd to cluster GCM vector
\item Shepherd to furthest agent from cluster vector
\item Cluster GCM to goal vector
\item Furthest agent from cluster to goal vector
\item Furthest agent from cluster to cluster GCM vector
\item Cluster GCM velocity vector
\item Furthest agent from cluster velocity vector
\item Number of agents in simulation
\end{enumerate}
    
The above inputs are computed from a series of raw observations:  Timestep number; Goal $X$ and $Y$ coordinates; Shepherd $X$ and $Y$ coordinates; Cluster GCM $X$ and $Y$ coordinates; Furthest agent from cluster $X$ and $Y$ coordinates.

\subsubsection{Teaching Plan and Lessons} 
The shepherding behaviour is deconstructed into switching, driving and collecting with each of the two latter behaviours individually trained on separate neural networks, and hence lessons targeted at these specific behaviours are required. 

The research from previous stages is consolidated and applied to give the following suggested scenarios within the training curriculum. It is worth noting here that the first sub-sub-level in the drive and collect sub-levels spawn the shepherd at the drive and collection points, respectively. This is based on the curriculum learning concepts explored in \cite{florensa2017} where the learner starts in a state close to the desired goal where it has a higher chance of achieving it, and then is moved further from that state as it progresses. This approach ensures that there is a higher percentage of ideal behaviour present in the training data where the shepherd is at the desired driving and collecting points respectively, than without it.

In the current obstacle-free task, the shepherd and sheep spawn within an unbounded environment (i.e. they will be free to move infinitely in all directions) simulating an infinite paddock. This is done to ensure the collection points and drive points can be navigated to freely, allowing the purist form of Str\"{o}mbom\textquoteright s model without interference from obstacles such as fence lines.

\begin{enumerate}[label*=\arabic*.]

\item \textit{Drive Behaviour:} The shepherd drives a tightly collected cluster to the goal location, with the simulation ending when this is achieved. The population of the cluster will systematically increase from 50 to 100 then 190 $\pm 10$ in each sub-sub-level based on the results of tests in the variable impact determination step.

\begin{enumerate}[label*=\arabic*.]

\item \textit{Straight drive lesson:} The shepherd spawns at the drive point i.e. directly behind the cluster relative to the goal, meaning it only needs to drive the cluster straight to the goal. 

\item \textit{Random drive lesson:} The spawn locations of the shepherd, goal and GCM are randomised within the environment.

\end{enumerate}

\item \textit{Collect Behaviour:} The shepherd collects at least one agent which spawns at a distance greater than $f(N)$ from the cluster. The simulation will end when the cluster is fully collected (i.e. all agents at a distance $<f(N)$ from GCM). The population of the cluster will systematically increase from 50 to 100 then 190 $\pm 10$ in each sub-sub-level based on the results of tests in the variable impact determination step.

\begin{enumerate}[label*=\arabic*.]

\item \textit{Straight collect lesson:} The shepherd spawns at the collection point; i.e. directly behind the furthest sheep relative to the cluster, meaning it only needs to herd the agent straight to the cluster.

\item \textit{Single agent random collect lesson:} The spawn locations of shepherd, goal, GCM and furthest sheep are randomised within the environment, with the furthest sheep radius from the GCM being randomised between $1.25f(N) \to 2f(N)$.

\item \textit{Spread cluster collect lesson:} All spawn locations are randomised with the cluster being spread over a distance greater than $f(N)$, leading to the shepherd needing to collect the entire population of agents.

\end{enumerate}

\item \textit{Open Environment Behaviour:} We also design three lessons to train the non-curriculum AIES, and for the testing and comparison of all AIES variances. For this reason these simulations end only when the overall task is complete (i.e. a collected herd is at the target). The three cases are made up of:

\begin{enumerate}[label*=\arabic*.]

\item \textit{Collected herd lesson:} A fully collected herd, target and shepherd spawn with random positions.

\item \textit{Single separated agent lesson: } A single agent spawns away from the herd, with all spawn positions randomised.

\item \textit{Spread herd lesson:} Cluster spread over a distance greater than $f(N)$, with all spawn positions randomised.

\end{enumerate}

\end{enumerate}

Simulations are broken up into groups of 30, to allow for 10 individual simulations at each cluster size. The samples are collected in small groups like this rather than all at once to allow for more modular data sets that can be mixed and matched during training as needed. Initially the break down of samples for the curriculum is to have four sets of thirty simulations in both the driving and collecting lessons, therefore in total 240 individual simulations are required. Within driving, one set will comprise of sub-level 1.1, and the remaining three will comprise of sub-level 1.2. This was chosen to allow for more samples for the more complex scenario which will be harder to learn, but a single set is expected to be sufficient to inform the ideal behaviour from 1.1. Similarly, collecting will be broken up into one set of 2.1, two sets of 2.2 and one set of 2.3. We have two sets in level 2.2 because of the importance of this behaviour. If the AI can learn to gather an individual agent, technically gathering an entire cluster will comprise of gathering each furthest agent individually.

\subsubsection{Evaluation} 
Analysis of the learning gradients found in the formative assessments on the quality and number of samples collected in each behaviour can be fed back into the curriculum design to make future improvements.

\subsection{Game Creation} 
Based on the findings of Curriculum design stage, a game with user input and data collection functionalities could be designed. The curriculum should be implemented into the game, with different levels being equivalent to the designed lessons.
    
The shepherd's directions is controlled by a human who is providing the target data. This was implemented using the mouse to provide a unit vector between the shepherds position and the shown mouse position, effectively having the shepherd follow the plotted mouse position each timestep. The mouse was chosen due to its ability to provide continuous directional data, as opposed to the arrow keys for example which would only provide limited discrete input.

\subsection{Data Collection and Analysis} 
A human (the first author) played 480 game simulations. The data was recorded to create approximately 200,000 training samples. To provide continuity in the behaviour of the AIES, stopping was replaced with a strategy whereby a circular motion away from the direction of travel was used. 

    
\section{Experiments}\label{experiments}
Two sets of experiments were run. Firstly, the summative assessment of the five fully trained AIES comprised of one curriculum based learner and four non-curriculum based learners. This was comprised of 100 simulations of our open environment, with pseudo random spawn positions so as to expose all five of the AIES to an identical test set. The data collected was for the purpose of comparing each of the AIES ability to exhibit the shepherding behaviour, within the minimum time-frame suggested in~\cite{strombomMann2014}. With the defined purpose of shepherding as driving a collected cluster of agents to a goal location, the minimum GCM distance to the goal achieved in each simulation was recorded as a metric of driving ability. Similarly the average distance of the furthest sheep to the GCM for each simulation was recorded as a metric of collecting ability. Finally a percentage of the binary success rate of each simulation was recorded as an overall shepherding competence metric. 
    
The Formative assessment experiments were run, comparing the same metrics as above but only on the single non-curriculum based AIES assessed as being the most comparable to the curriculum based learner. The purpose of this was to observe and compare the learning efficiency of each of the AIES with increasing training sample sizes. This was measured by finding the learning gradient of the AIES in each of the above metrics.
    
\subsection{Summative Assessment Experiments}

To determine the effectiveness of the curriculum based AIES, it was compared against a number of non-curriculum based AIES. This was used to confirm that the non-curriculum based learners are not at a disadvantage to the curriculum based learner, ensuring the only variable was learning via the curriculum. The curriculum based AIES was comprised of two neural networks (one for driving and one for collecting), each containing 10 hidden nodes in a single hidden layer and 53,890 training samples. Two of the non-curriculum neural nets contained 20 hidden nodes, matching the total number of hidden nodes in the curriculum neural nets, and two contained 10 hidden nodes, matching the number of hidden nodes in each of the curriculum AIES individual neural nets. Another parameter that needed to be tested for was the training sample size. The curriculum AIES's training scenarios were on average smaller than the open environment that the non-curriculum AIES was trained on. For this reason, two of the AIES were trained on a sample size 54,629 matching that of the curriculum AIES (consequently, containing less full training simulations), and two were trained with a larger sample size of 107,986 to match the total number of full training simulations the curriculum AIES had access to. 

%

\subsection{Formative assessment experiments}

In this experiment, the curriculum-based shepherd and the best non-curriculum variant had their learning gradients compared by assessing our metrics while varying training sample size. The purpose of this experiment was to show the efficiency of learning with the hypothesis that the curriculum based learner would learn at a greater rate than the non-curriculum based learner.
    
\section{Results and Analysis} \label{results} 

\begin{figure*}[ht]
\centering
\begin{subfigure}{0.3\textwidth}
    \centering
\includegraphics[width=\linewidth]{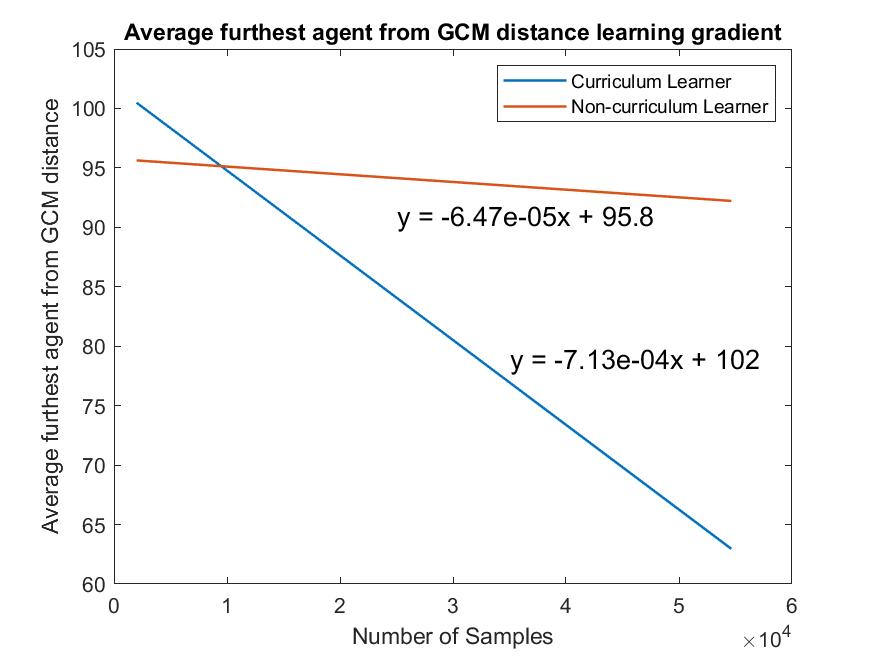}
    \caption{Average GCM to furthest agent distance at increasing sample numbers}
    \label{fig:avgGCM2FurthPlot}
\end{subfigure}
\begin{subfigure}{0.3\textwidth}
\centering
\includegraphics[width=\linewidth]{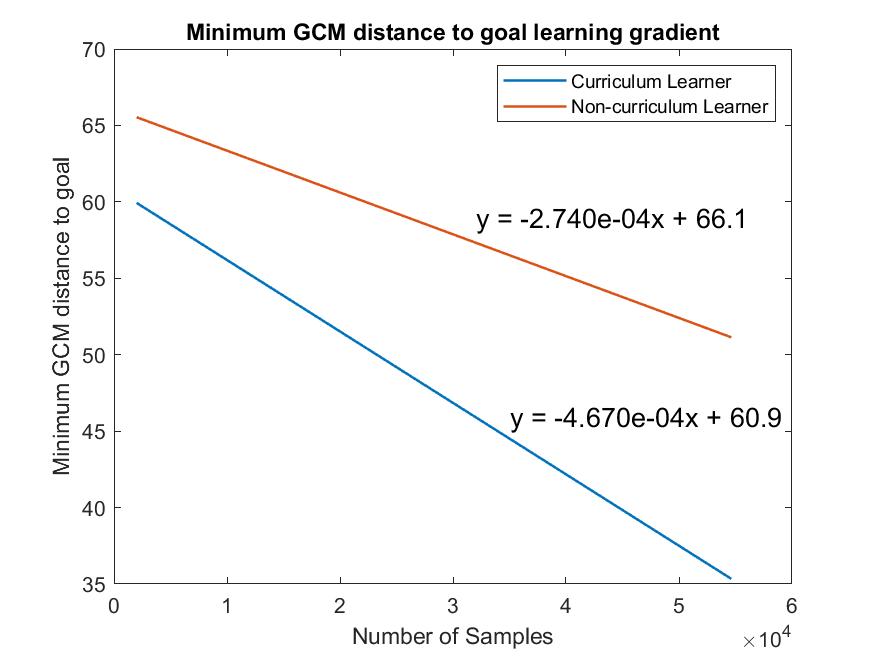}
    \caption{Average minimum GCM to goal distance at increasing sample numbers}
    \label{fig:minGCM2GoalPlot}
\end{subfigure}
\begin{subfigure}{0.3\textwidth}
    \centering
\includegraphics[width=\linewidth]{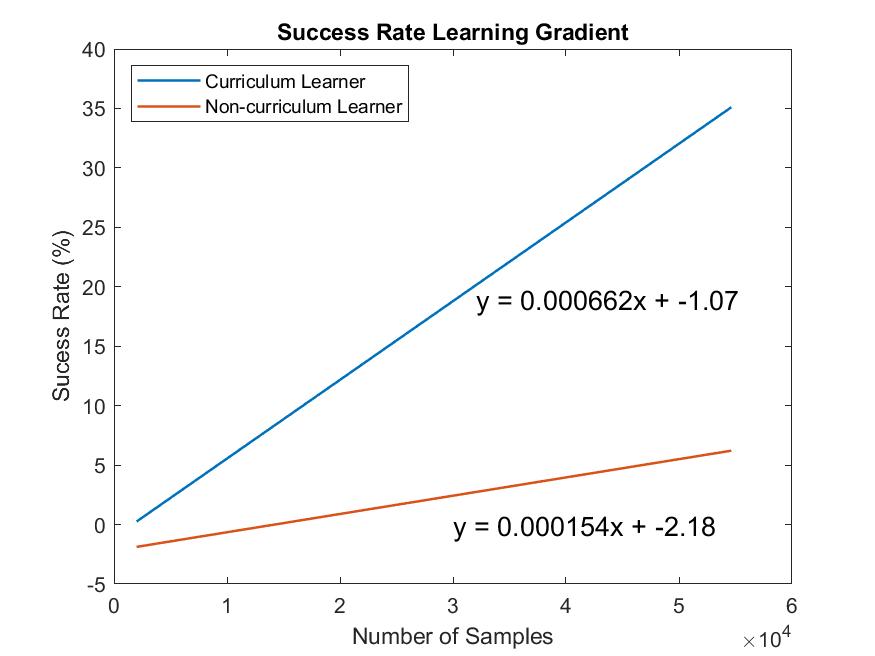}
    \caption{Success rate at increasing sample numbers}
    \label{fig:successRatePlot}
\end{subfigure}
\caption{Formative assessment results.}
\label{fig:progressAssessmentPlots}
\end{figure*}

Table~\ref{tab:neuralNetResults} shows the results from the summative assessment experiment described above. Figure~\ref{fig:progressAssessmentPlots} shows the learning gradient of success rate, average furthest agent from GCM distance and the minimum GCM distance to goal distance. A high positive gradient for success rate and a high negative gradient for furthest agent and goal to GCM distances indicate that curriculum learning is more effective.

\begin{table*}[h]                                                         \centering                                                      
\begin{tabular}{x{1.1cm}x{1cm}x{1.2cm}x{1.5cm}x{1cm}x{1.7cm}x{1.5cm}x{0.9cm}x{1.5cm}x{1.5cm}}\hline
Shepherd ID No.     &  Mean min GCM to goal distance    & Standard deviation min GCM to goal distance   & Driving improvement statistical significance  & Mean furthest sheep to GCM distance & Standard deviation mean furthest sheep to GCM distance & Collecting improvement statistical significance & Task success rate & Success rate improvement standard deviation & Success rate statistical significance\\ \hline
C1          & 36.16 & 37.91 & N/A           & 64.39     & 33.15     & N/A       & 32\%  & 4.66  & N/A       \\ 
NC1         & 52.33 & 38.98 & 0.3372        & 90.81     & 35.89     & 0.2296    & 7\%   & 2.55  & 0.000     \\ 
NC2         & 38.75 & 38.27 & 0.4721        & 83.60     & 36.43     & 0.2981    & 22\%  & 4.14  & 0.1131    \\ 
NC3         & 63.22 & 39.75 & 0.2483        & 90.75     & 28.36     & 0.1762    & 2\%   & 1.40  & 0.000     \\ 
NC4         & 54.21 & 39.78 & 0.3264        & 87.49     & 25.53     & 0.1841    & 7\%   & 2.55  & 0.000     \\ \hline
\end{tabular}
\caption{Summative assessment results} 
\label{tab:neuralNetResults}
\end{table*}

The results shown in Table~\ref{tab:neuralNetResults} firstly confirm that the curriculum learner did not have any advantages in terms of sample size or node number over the baseline AIES. This is shown by the fact that even the best performing non-curriculum AIES, NC2, did not perform better than C1 (the statistical significance of difference was insignificant) even with double the number of training samples. Assessment on NC3 and NC4 were conducted in order to confirm that the curriculum learner did not have the benefit of a smaller number of hidden nodes but these still perform poorly when compared to NC1 and NC2, confirming that this is not the case.
        
There is an improvement seen in the curriculum learner\textquoteright s ability to drive the herd towards the target, observable through the mean min GCM to goal distances. As is shown in Table~\ref{tab:neuralNetResults}, C1 has an average minimum simulation distance of 36.16 units, whereas NC1 has a distance of 52.33 units. Smaller values are better. This is an improvement of approximately 44\%. Similarly, C1\textquoteright s ability to collect the herd is improved, observable in the mean furthest sheep to GCM distance. C1\textquoteright s average distance is 64.39, whereas NC1\textquoteright s distance is 90.81, and again a small value here is better, as it means that the shepherd has been able to keep the cluster more tightly collected. This is an improvement of approximately 41\%. However, due to the relatively large standard deviation of these values, at a confidence level of $\alpha = 0.01$ the improvements are not statistically significant. For collecting and driving the improvement p-value is found to be $p_c = 0.23$ and $p_d = 0.34$ respectively, thus we must in this isolated case accept the null hypothesis that the curriculum has had no impact on the shepherd\textquoteright s ability to collect and drive individually, based on these metrics. 

Interesting to note however, is the relatively significant improvement in success rate. Table~\ref{tab:neuralNetResults} shows that C1 had a success rate of 32\%, compared to NC1's 7\%. This is an approximate 457\% improvement, and at a negligible p-value we can reject the null hypothesis and assume that the curriculum has had some impact on the shepherd's ability to complete the overall task of shepherding. 

Figure~\ref{fig:progressAssessmentPlots} shows the pattern of improvement for the curriculum and non-curriculum learners, when provided with an increasing number of training samples. As discussed the success rate improvement was statistically significant, and as we can see in Figure~\ref{fig:successRatePlot} there is an observable pattern of faster improvement in the curriculum AIES over the non-curriculum agent. While the individual results at each sample size have been statistically insignificant in the collecting and driving metrics, the same pattern can still be observed and is worth discussing.

Figure~\ref{fig:avgGCM2FurthPlot} shows our collecting metric learning plots. From the best fit equations we can see a curriculum to non-curriculum learning rate ratio for collecting of:
    \begin{align}
        M_C = \frac{m_{C1}}{m_{NC1}}    & = \frac{-7.13e-04}{-6.47e-05} \approx 11.02
    \end{align}
The curriculum learning rate is larger than that of the non-curriculum. It is worth noting that the larger constant in the best fit equation for the curriculum learner ($102 > 95.8$) suggests that the non-curriculum learner is initially more competent at collecting based on this metric. However this quickly changes with the fewer samples of 10,000, evident by the point of intersection at this value. This means that as hypothesised, the curriculum is allowing C1 to learn more efficiently than NC1 in terms of collecting, based on this metric.
    
Secondly, Figure~\ref{fig:minGCM2GoalPlot} shows our driving metric learning plots. From the best fit equations, we can see a curriculum to non-curriculum learning rate ratio for driving of:
    \begin{align}
        M_D = \frac{m_{C1}}{m_{NC1}}    & = \frac{-4.67e-04}{-2.47e-04} \approx 1.70
    \end{align}
This again confirms that the curriculum is allowing C1 to learn to drive at a faster rate than NC1, based on this metric. The smaller constant in the line of best fit ($60.9 < 66.1$) also suggests that C1 is more competent initially, and continues this by learning at a faster rate.
    
Finally, Figure~\ref{fig:successRatePlot} shows our overall task competence learning rate. From the best fit equations we can see a curriculum to non-curriculum learning rate ratio for overall competence of:
\begin{align}
M_S = \frac{m_{C1}}{m_{NC1}} & = \frac{6.62e-04}{1.54e-04} \approx 4.30
\end{align}
This signifies that C1 is able to learn overall task competence at a higher rate than NC1 as hypothesised. The larger constant in the line of best fit ($-1.07 > -2.18$) also tells us that C1 is more competent initially than NC1, and continues this by learning at a faster rate.

\section{Conclusion and Future Work}\label{conclusion}

Solving complex problems through supervised learning requires large amounts of accurately labelled data. The shepherding problem is one such example due to its large state space. Collecting this data through means of experiments with sheepdogs for example is a large scale commitment and would need to be reimplemented every time further development of the model is desired. We suggest that a solution to this is the development of a curriculum to efficiently create higher quality training samples and data collection through interactive simulations. This allows for natural deconstruction of behaviours, providing benefits such as higher explainability, modularity and maintainability in the lower level learned behaviours.
    
Our results demonstrate the benefit in efficiency and effectiveness of the suggested AI Education Framework with human interaction to generate training data. To demonstrate this, a curriculum for a simple case of the shepherding problem was developed and implemented in the form of a shepherding game which collected the required data from human demonstration, to be presented as training samples to the supervised learning AIES neural nets. We demonstrated that the curriculum based AIES was able to learn more efficiently and effectively than the baseline non-curriculum learner, and claim that this is due to a combination between the deconstructed behaviour implementation, and the higher quality training samples provided through the developed curriculum.

For future work, as detailed, in our implementation the curriculum learner's switching behaviour was hard-coded in an if statement. As the number of behaviours increase, it makes sense to train a neural network to switch between different behaviours. Moreover, our curriculum caters to a simple shepherding case, where the paddock is infinite, contains no obstacles or gates and the sheep and shepherd characteristics do not change. Implementing further levels to handle more complex cases is necessary for the behaviour to be applied to the more general case.

\bibliographystyle{IEEEtranS}


\end{document}